# Open-source software implications in the competitive mobile platforms market.

Salman Q. Mian[1], Jose Teixeira[2], Eija Koskivaara[3]

[1]Nokia Siemens Networks (NSN), Linnoitustie 6,
02600 Espoo, Finland
[2]Turku Center for Computer Science (TUCS), Joukahaisenkatu 3-5 B,
20520 Turku, Finland
[3]Turku Shool of Economics (TSE), Rehtorinpellonkatu 3,
20500 Turku, Finland

{Salman.Mian}@uta.fi
{Jose.Teixeira, Eija.Koskivaara}@tse.fi

**Abstract.** The era of the PC platform left a legacy of competitive strategies for the future technologies to follow. However, this notion became more complicated, once the future grew out to be a present with huge bundle of innovative technologies, Internet capabilities, communication possibilities, and ease in life. A major step of moving from a product phone to a smart phone, eventually to a mobile device has created a new industry with humongous potential for further developments. The current mobile platform market is witnessing a platforms-war with big players such as Apple, Google, Nokia and Microsoft in a major role. An important aspect of today's mobile platform market is the contributions made through open source initiatives which promotes innovation. This paper gives an insight into the open-source software strategies of the leading players and its implications on the market. It first gives a precise overview of the past leading to the current mobile platform market share state. Then it briefs about the open-source software components used and released by Apple, Google and Nokia platforms, leading to their mobile platform strategies with regard to open source. Finally, the paper assesses the situation from the point of view of communities of software developers complementing each platform. The authors identified relevant implications of the open-source phenomenon in the mobile-industry.

**Keywords: Open-source, Platform-strategies, Mobile-industry, Mobile-platforms, iOS, Android, Symbian, Maemo**

# 1 Introduction

The open-source software phenomenon continues, persistently capturing the attention of both scholars and practitioners. It started in 1985, when Richard Stallman founded the Free Software Foundation promoting the idea of freedom in software. The Foundation, still very active today, promotes that software could run freely and that correspondent software source code could be studied, changed, copied, published and also distributed freely.

Raymond (2001) popularized the phenomenon by studying the development of the first free operating system known as GNU/Linux. It was claimed that Linus Torvalds steered a totally new way of developing software by making use of thousands of volunteer developers collaborating over Internet in a distributed “organization” towards a common goal. Representing a more efficient way than the traditional hierarchical and controlled way used by corporate software houses.

The echo of open-source software development attracted an interest particularly from the economic scholars. Lerner and Tirole (2001) made a preliminary exploration of the economics of open-source by assessing the extent to which economics literature on “labor economics” and “industrial organization theory” could explain the open-source phenomenon. The mentioned research brought some answers on what motivates open-source developers, compared the different programming incentives between open-source and proprietary settings, and highlighted the favorable organizational and governance characteristics for open-source production.

Many other relevant contributions followed Tirole’s research agenda: such as Paajanen (2007), that developed a multiple case study on the licensing for open-source software; and Bonaccorsi and Rossi (2003) that discussed the coexistence of open-source and proprietary software in organizations. Other studies give evidence to a large scale adoption of open-source software by organizations; perhaps one of the most clear evidences illustrating the growing economic impact of open-source comes from Finland. According to Helander et al. (2008), 75% of the studied Finnish firms were using open-source software, a enormous increase since only approximately 13% of the same firms were using open-source software according to an analogous survey conducted in 2000.

The term platform is conceptually abstract and widely used across many fields. Within this research, the term platform maps the concept of computer-based platform as in Morris and Ferguson (1999), Bresnahan and Greenstein (1999) and West (2003). As mentioned by West (2003), a platform consists of an architecture of related standards, controlled by one or more sponsoring firms. The architectural standards typically encompass a processor, operating system (OS), associated peripherals, middle-ware, applications, etc.

Platforms can be seen as systems of technologies that combine core components with complementary products and services habitually made by a variety of firms (complementers). The platform leader and its complementers jointly form an “ecosystem” for innovation, that increases platform's value and consequently user’s adoption (Gawer and Cusumano 2008). For instance, the current leaders of the video games industry, operate by developing the hardware consoles and its peripherals while providing a programmable software platform that allows complementers to develop games on top of their systems. Attracting more game developers to the platform means more and much better games, an increase in value for the end users (video game players).

It is empirically observable that the current mobile industry is shifting from a products-war paradigm to a distinct platforms-war situation. Rather than vendors competing on a basis of its perceived product features, we observe a complex network of vendors under the umbrella of a platform aimed in becoming the “defacto” standard under a market subject to the forces of network externalities described by Shapiro and Varian (1999). This empowers the importance of platform complementers such as telecommunication operators, semi-conductors components makers, producers of hardware accessories, 3rd party software application developers, etc. As mentioned by West (2003), device makers differentiation might not be driven by higher architectural layers but by efforts in the systems integration and design.

The future of the mobile platform market is as debatable as global warming issues, though the earlier inevitably presents a promising and better future. Just a look at the past ten years give an impression of a primitive era, which has grown out to be a huge bundle of innovative technologies, Internet capabilities, communication possibilities, and ease in life. It all started from a simple product phone, then to a smart phone but the technological developments forced the use of a more general term, such as mobile device. In the current mobile devices market, there are still doubts over what defines a successful mobile device, is it the hardware, or the operating system, or the ease of use. The success of a mobile device has become a combination of all these things including the brand name, open source code, variety of applications, and many more aspects. According to Eric Schmidt, Google’s executive chairman, brand is just a little part, the high-tech war between companies is about coming up with the ideas, revolution in applications, and the chance to lead in the future mobile devices industry (Zakaria, 2011). From the consumer’s point of view, the mobile devices market is moving very fast and is volatile. In a more general perspective, it has become a race to be the best operating system for the Internet enabled phones (Ocock, 2010).

The current mobile device platforms market is mainly dominated by Google, Apple, and the one with its own legacy, Nokia. When it comes to platform, each of them promotes their own, Android from Google, iOS from Apple and the historical Symbian from Nokia. While Apple’s iOS is restricted to their own device iPhone,

Symbian is mostly to Nokia, Motorola and Samsung phones especially the current version Symbian^3. Android from Google is the most versatile in terms of adoption, being used in mobile devices from other companies such as HTC, Samsung, and LG, to name a few. This research mainly focuses on Android, Symbian, Maemo and iOS platforms leaving the others like RIM, Windows mobile platform etc. This is mainly because other platforms either do not have a significant market share or they clearly state that their interests do not match with that of the open source community.

Symbian retains the credit of being the oldest smart-phone platform in use and it corresponds to them being the biggest operating system by market share at the moment (Ocock, 2010). All of these platforms have market share varying according to geographical locations in the world. But in general others have captured the market share mostly from Symbian in the last few years. The Fig. 1 below shows the worldwide smart-phone market shares percentages by operating systems in the last two years, according to Gartner, 2011.

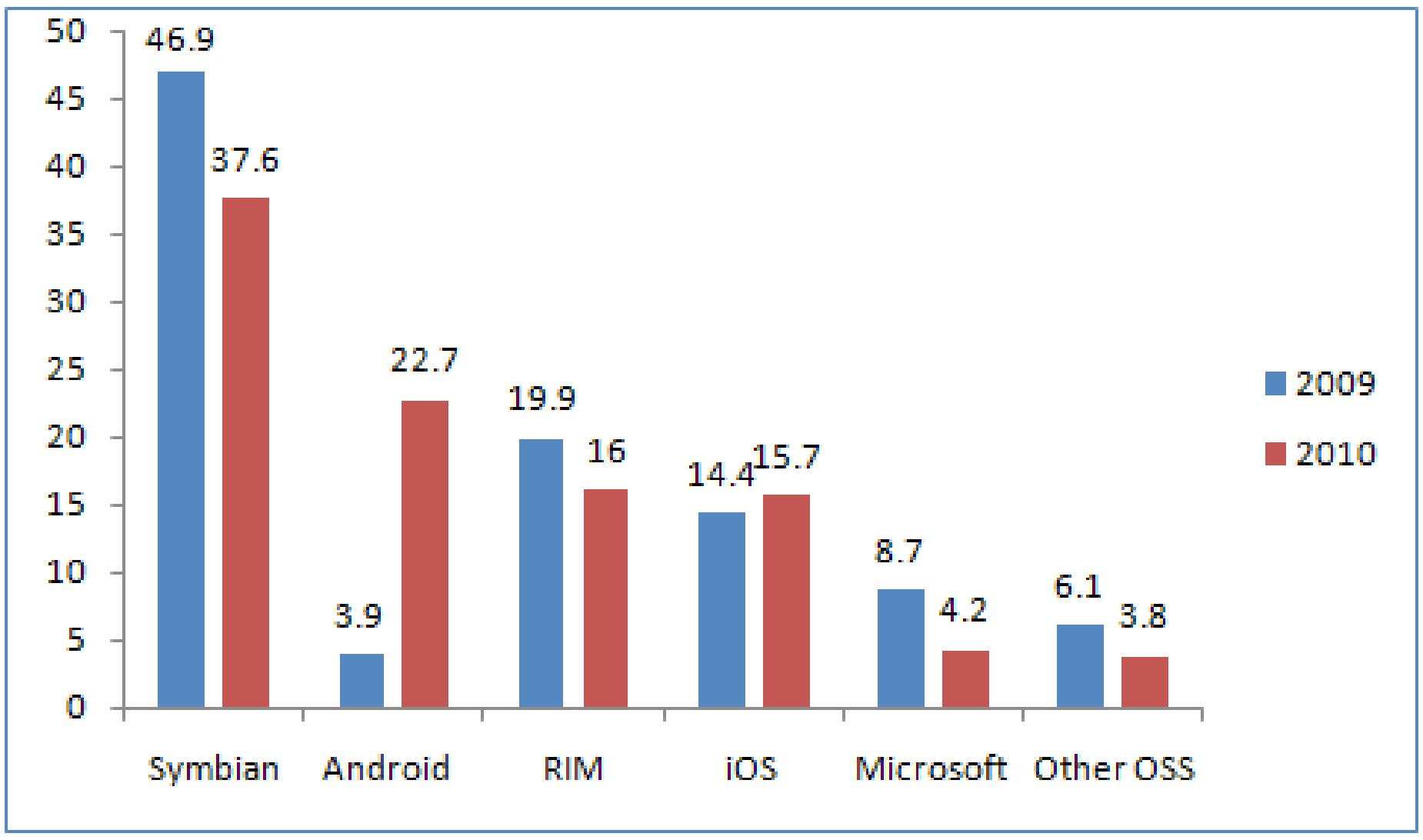

Fig. 1.Worldwide smart-phone Market shares (%) by platform in 2009/2010 (Gartner, 2011)

The technological developments in various mobile device platforms has eventually introduced tough competition, with eventually consumer winning in the end. One of the adoptions on its way is Microsoft's Windows phone 7 OS taken by Nokia, which is a strategic step taken by the company assessing the current market (Nokia press release, 2011). However, with increasing competition, the mobile devices industry has also been marred with lawsuits. In the recent years, the above mentioned supreme leaders have now and then been involved in various patents and copyright cases against each other. Another aspect of the current platforms market is the code being open source (meaning available to everybody), with the perception of achieving innovation and creativity by getting all the developers involved. However, among the above companies this positive initiative varies on different grounds.

On the other hand, an Open Handset Alliance led by Google was founded in Nov, 2007 with the purpose of accelerating innovation in mobile and to richly improve the consumer experience. The alliance is a group of 84 technology and mobile companies which together released the Android with the aim of deploying handsets and services using the Android platform. The alliance is committed to great openness for the development of the Android platform through open software and applications (Open Handset Alliance, 2011).

Considerable research was established on technological platform strategies, being briefly identified here: Anchordoguy (1989) exploited the rich competition between computer platforms in Japan while the western world was being monopolized by IBM. Bresnahan and Greenstein (1999) examined thirty years of the computer industry from a pure economical perspective. West (2003) investigated in detail, the hybrid strategies from PC vendors that attempted to combine the advantages of open-source software while keeping tight control and differentiation in its platforms. The Japanese PC and Gaming industry carefully reviewed by Hagiu (2004) introduced the concept of multi-sided platforms and surveying prevalent business models adopted by dominant platform makers. Gawer and Cusumano (2008) made key contributions by introducing the coring and tipping two strategic abstractions. However, with a stronger empirical relevance of open-source factors and with a completely new mobile industry in context, current established research needs to be further developed.

## 2 Methodology

As the title of this paper suggests, the research question of this study is “What are the open-source software implications in the competitive mobile platforms market?”. Forced by the magnitude of the research question, the authors fractionated the research problem with the following three research questions: First, “What are the OSS components integrated by Apple, Google and Nokia in their mobile platforms?”; Second, “What are the open-source platform-based strategies employed by Apple, Google and Nokia?”; And finally, “How are the 3rd party developers coping with the announced strategies”?

The authors addressed the research questions with Yin (1991) case study research methodology. The research authors had tiny or no control over networked behavioral events within the complex market of mobile device platforms being studied. Moreover, this research focuses on contemporary phenomenon in a real-life context where the boundaries between phenomenon being studied and its context is not obvious. According Flyvbjerg (2006), an in-depth case study research approach provides a systematic way of looking at events, collecting data, analyzing information, and reporting valuable results as knowledge.

Following Yin (1991) case study methodology our case is “implications of the open-source software in the mobile industry” and our units of analysis are the four mobile platforms developed by the Apple, Google and Nokia, giants of the telecommunication industry: “iOS”, “Android”, “Symbian” and “Maemo”. According to the same author’s taxonomy, our research is a multiple descriptive case study. It follows an embedded design with multiple units of analysis where the authors are looking for consistent patterns of evidence across the four units within the same phenomenon being studied.

The mobile devices industry is recent, and the emergence of mobile devices platforms is ever newer. Perhaps the EPOC system, that surged in mid 1996 empowering Psion devices, is the first mobile device platform. Later known as Symbian, it provided differentiated structures for handset device makers and application developers, instantiating the concept of multi-sides platforms described by Hagiu (2005).

The novelty of the phenomenon being studied constrained the work for the authors in finding early theoretical knowledge addressing the research problem. The researchers strongly believe that it is dangerous to generalize research from the previous two decades on the computer-based platforms, to mobile platforms: the market players are different, the technology and geographies too. With the lack of early knowledge the authors were forced to seek an exploratory case-study over descriptive or explanatory approaches. The authors share the same view as Flyvbjerg (2006), crediting that concrete and context-dependent knowledge is often more valuable that vain search for predictive theories and universals.

The data was systematically collected between the 20th of May and 8th of April 2011, from a set of Internet sites which are mostly publicly available. One of the research authors subscribed the software development programs from Apple and Nokia to get access to information targeting each platform’s third party software developers. The following Table 1 presents the different websites from where the case study data was collected. The authors explored a key strength of the case study method by making use of multiple sources and techniques (Yin 2002). The systematically analyzed Internet sites greatly differ in vendor control, editorial constraints, pluralism and interactivity. The data sources can be categorized in four group-types with regard to developers goals and policies: First, press releases provided by the platform vendors were studied; followed by software development portals and discussion forums covering each of the platform, then generalist business, economical and technological press; and finally, websites with a very strong focus on reviewing the personal electronics industry.

**Table 1.** Internet captured research data sources description

| Site | Description | Vendor controlled ? | Confidentiality |
|---|---|---|---|
| http://www.apple.com/pr/ | Apple press release | Yes | Public |
| http://www.google.com/press/ | Google press release | Yes | Public |
| http://press.nokia.com/ | Nokia press release | Yes | Public |
| http://developer.apple.com/devcenter/ios | Apple developers portal | Yes. Partially | Public or constrained to 3rd party developers |
| http://developer.android.com/resources/community-groups.html | Android developers community groups | Yes. Partially | Public |
| http://www.forum.nokia.com/ | Nokia interface with its developers | Yes. Partially | Public or constrained to 3rd party developers |
| http://maemo.org/ | Nokia Maemo platform community portoal | No | Public |
| http://symbian.nokia.com/ | Nokia Symbian platform portal | Yes. | Public or constrained to 3rd party developers |
| http://www.businessweek.com | A weekly business magazine. With strong USA focus. | No | Public |
| http://www.wired.com/ | Magazine covering how technology affects culture, the economy, and politics. Strong USA focus | No | Public |
| http://news.cnet.com/ | CNET provides reviews of both consumer electronics and software. | No | Public |
| http://thsnews.com | Independent generalist news provider. | No | Public |
| http://ostatic.com/ | Portal reviewing open-source software and | No | Public |
| | services. | | |
| http://globalpublicsquare.blogs.cnn.com/ | Generalist blog feed by TIME and CNN journalists among other contributors. | No | Public |
| http://www.zdnet.com/ | Technology news and product reviews | No | Public |
| http://slashdot.org/ | Technology news forum hosting many discussion on open-source topics | No | Public |
| http://www.engadget.com | A web magazine with focus on consumer electronics. With strong USA focus. | No | Public |
| http://eu.techcrunch.com/ | A blog based edition covering Web 2.0 and Mobile start-ups. EU focus. | No | Public |

The data was collected and classified by taking into account the Romano et al. (2003) methodology for analyzing web-based qualitative data encompassing the three, elicitation, reduction and visualization processes. Data elicitation and reduction was performed over the Internet by collaborative manners, using the popular googledocs web-based software. Reduction included the grouping of data into different codes derived from previous theory and observed text data, such coding took into account some of the qualitative research principles stipulated by Seaman (1999). Both the popular office-suite package and the freeMind software application were intensively used during the visualization phases, allowing the authors to identify patterns and structures which permitted drawing of conclusions.

## 3 Findings and research agenda

Targeting the first research question, assessing what open-source software components are integrated by Apple, Google and Nokia in their mobile platforms, the researchers identified most of them from the official platform websites. These are enumerated in the following Table 2, in terms of the software packages resulting from the vendor integration of technological assets provided by the open-source community. Every package found was carefully investigated by the authors and rich data was collected for post-analysis. Collected evidence included, a brief description of each package, the correspondent open-source project website, the open-source license in which the software is distributed and, wherever possible, the organization behind the identified software project.

**Table 2.** Platforms architecture reviewed data sources

| Platform | Website |
|---|---|
| iOS 4.3.3 | http://www.opensource.apple.com/release/ios-433 |
| iOS 4.3.3 | http://www.opensource.apple.com/release/developer-tools-40 |
| Android 2.3 | http://android.git.kernel.org/platform/external/ |
| Maemo 4.1 | https://garage.maemo.org/docman/view.php/106/354/maemopackages-20080725.ods |
| Maemo 4.1 | http://maemo.org/maemo_release_documentation/maemo4.1.x/ |

The software packages obtained from the list of websites in Table 2 were verified to present real world devices. By this, the researchers ensured that each package found was present in a physical device empowered by the studied platforms. However, due to technical and legal constraints, it was only possible to do such verification on the Maemo and Android platforms, using a Samsung Galaxy S mobile phone and a Nokia N810 internet tablet. All packages found were present in the real world physical devices.

A considerable amount of open-source software packages were found, 28 packages within the iOS 4.3.3 platform, 108 packages within the Android 2.3 platform and 151 packages in the Maemo platform. During the data collection period, on 6th of April, Nokia announced a radical change on the organization strategy abandoning the Symbian platform’s open-source strategy and the corresponding Symbian foundation. The source-code is no longer available, and the official Nokia platform website at http://symbian.nokia.com read “Not open-source, just open for business”. And the platform website is not even available from non-official websites like http://sourceforge.net/projects/symbiandump. Since the Symbian operating systems did not had strong connections with the open-source community from the start, the authors decided to drop any investigation regarding what Symbian platform packages were derived from the open-source community.

However it is important to note especially when addressing the second research question, that the last version of the Symbian platform included the open-source Qt technology. Described at http://qt.nokia.com, it allowed third party application developers to target different Nokia platforms, using a single software development kit. Even if Maemo version 4.1 was able to run applications developed in Qt, official support of Qt was only announced for the Maemo version 5. Nokia purchased Trolltech, the Norwegian company behind Qt, in January 2008 for $153 million and

immediately started promoting a three platforms strategy, which included Symbian S40, Symbian S60 and Maemo, targeting different ranges of devices. But a unified software development kit based on Qt allowed developers to quickly port their applications for different Nokia platforms and devices as well.

Much before the full implementation of the above mentioned three platforms strategy, the appointment of Stephen Elop as CEO on 21 September 2010 brought radical changes in Nokia strategy. The new CEO appointed the Windows Mobile platform as the new primary platform for Nokia, turning Microsoft from rival competitor to a strategic partner. On a curious note, Stephen Elop had moved to Nokia from a lead executive position at Microsoft. On February 2011, Nokia announced that the new Nokia Windows mobile platform will not support the Qt technology. This move caused strong dissatisfaction among Nokia third party software developers that they would need to completely re-develop their applications to a new platform with unknown capabilities. The open-source community behind Qt, with strong contributions from Bogdan Vatra quickly announced the port of the Qt core technology to the rival platform Android on March 2011. The link of Nokia with the Qt technologies ended in the same month with the partial sale of Nokia's Qt business operations to Digia, a Finnish IT services provider.

From the analysis of software packages resulting from the vendor integration of technological assets provided by the open-source community, we can conclude that Apple uses open-source software at low extend. iOS 4.3.3 makes extensive use of open-source components, but only in its operating system core, a layer completely hidden from the iPhone device users. The only exception is the web browsing technology which is based on the webKit open-source project available at http://webkit.org/. Apple is the main contributor to the webKit open-source project initiated as a fork of the KHTML open-source project from the KDE open-source community. It was noticed that Apple integrates older versions of open-source projects releases, perhaps seeking architectural simplicity and stability over the integration of the last project features, more prone to bugs.

Regarding open-source components integrated by the Android 2.3 platform, it is visible that Google uses open-source components at large. We can induce that the Android platform uses many recent versions of open-source project releases and that many of the used open-source components are heavily modified to facilitate adaptation to the platform operable architecture. Google and the Open Handset Alliance make use of a virtual machine and language interpretation technologies. Their aim is to try to keep the GPL license domain of the integrated software packages outside the developer software development interfaces, most probably to avoid possible legal litigations.

The Maemo 4 platform, in similitude with the Android platform, also integrates a large number of open-source components. However, it seems that Maemo architects are more satisfied with the original work from the open-source communities and therefore are not so intensively modifying the source code from integrated open-source projects. Moreover, Nokia allows its third party developers to directly access the integrated components programming interfaces. Thanks to the transparency of the Maemo open-source platform community, the following Table 3, illustrates how many packages are modified or directly integrated from the open-source projects. It is important to note that the number of packages are not correlated with the effort spent on its development, since each integrated project differs in size and complexity.

**Table 3.** Core decisions taken by the Maemo architecture and integration teams.

| Architecture Integration approach | Number of packages |
|---|---|
| OSS directly integrated from communities (upstream) | 68 |
| OSS modified by Nokia during the integration process | 79 |
| OSS project initiated and developed under Nokia umbrella | 49 |
| Closed source components developed under Nokia umbrella | 92 |
| Closed source components by 3rd parties | 2 |

One of the most interesting findings of the research is a list of open-source technology components, which are commonly integrated by the vendor platforms studied. Illustrating which open-source components are commonly used by Apple, Google and Nokia the following Fig. 3 reveals high similitude between the Android and Maemo platforms.

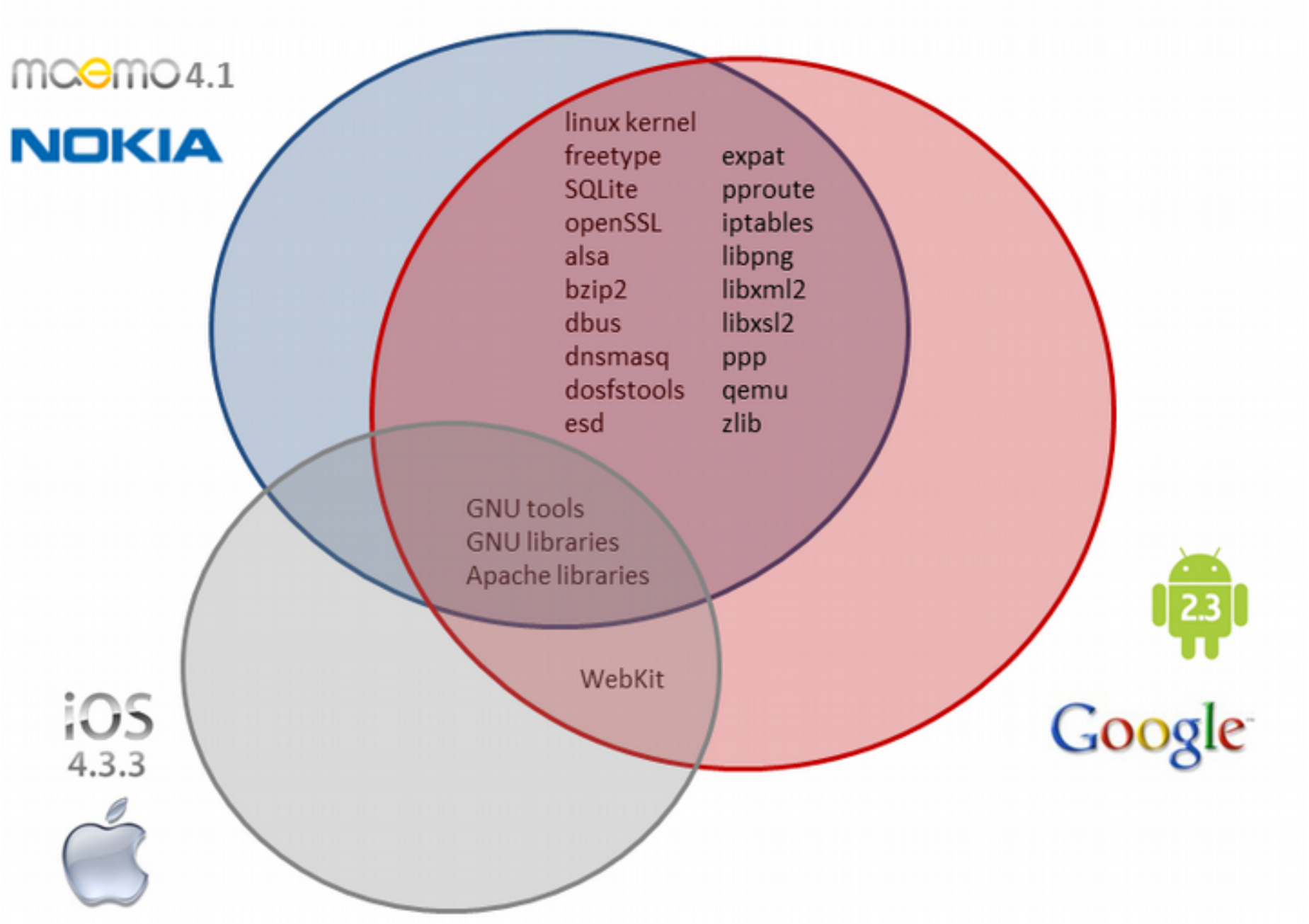

Fig. 2: Venn diagram with open-source technology commonly used by the three platforms.

Open-source mature tools and libraries from the long established and reputed GNU and Apache communities are integrated across the three platforms. Google Android and Nokia Maemo integrate many common open-source components in their architectures, while Apple seeks a distinct architecture. Apple IOS and Google Android use the same open-source browsing technology webKit. Curiously, webKit is a fork of the open-source project KHTML imitated by Lars Knoll and George Staikos in 2006 within the Trolltech/Qt sphere, and that Trolltech was acquired by Nokia in the first semester of 2008.

If vendors differ on what components they integrate, they differentiate even more on how they do it. Apple does not seem to be working “up-stream” with the communities of the open-source software. There are strong evidences that Google and Nokia work with and contract members from the open-source community. Both also try to integrate their own code modification/contributions to the source project. Forced by architecture decisions and dissatisfaction with the work produced by the open-source community, Apple and Google actively modify the packages they integrate. Nokia does it case by case, perhaps because Maemo architecture maps a more pure Linux-PC architecture from where most of the open-source software contributions come from.

From the sharing perspective, the platforms completely differ on how their core blue-prints are shared as public domain with the open-source community. Apple seems to provide the source code of core components in its platform to avoid legal litigation. Nokia provides circa 80% of its Maemo platform, but hides components like hardware adaptation, network connectivity and UI elements like sounds and fonts. Google provides almost 100% of the platform source code, however, it delays the release of the source code for selected versions. This means that Google wants to have a momentum where it protects the blueprint of its latest developments, for example it delayed the release of the Android 3.0 Honeycomb (a tablet-oriented version) for few months, keeping its source code out of the public domain. It is important to note that the development does not follow the Bazaar model presented by Raymond (2001), both Nokia and Google work with the open-source community over the Internet but under very tight control. There are strong evidences that both vendors maintain the repositories with a public version of their platform and keep a closed internal one as well.

Directly addressing the second research question, seeking a description of the open-source platform-based strategies employed by Apple, Google and Nokia, the following Table 4 captures key milestones employed by the vendors in concern. The presented information aims to provide a comprehensive detail of the different tactics in a platforms-war scenario that dominates the mobile-industry:

**Table 4.** Open-source mobile-industry milestones 2005-2011

| Year | Apple | Nokia | Google |
|---|---|---|---|
| | | Maemo.org goes online | |
| | | | Acquires Android Inc. |
| 2005 | | Nokia 770 Internet Tablet Starts Shipping, OSS phone based on Maemo | |
| 2006 | | | Buys YouTube |
| 2007 | Reinvents the Phone as iPhone | | |
| | | Buys Navteq, the maker of digital mapping and navigational software | |
| | | | Announced android, Founded Open Handset Alliance |
| 2008 | | Achieves 40% phone market share | |
| | | Acquires Trolltech (owner QT technologies) | |
| | | | Android was run on Nokia N810 |
| | | | Started Google I/O (Innovation in Open) annual conference |
| | Introduces the New iPhone 3G | Announces to make Symbian open source | |
| | | | Announed the on-deck Open Content Distribution system where developers can sell their application direct to the users. |
| | Apple releases iTunes version 8 | | World's first Android powered |
| | | | phone announced |
| | | | Made the entire Andriod source code, open source under the apache license; Launched G1 with HTC, first Android OS based mobile device |
| | | Nokia acquires Symbian Ltd | |
| | | | Introduces Google Latitude for mobile devices |
| | | | Samsung became the first among the top mobile manufacturers to adopt Android platform |
| | | Ovi Store by Nokia is available globally to an estimated 50 million Nokia device owners | |
| | Announces the New iPhone 3GS—The Fastest, Most Powerful iPhone Yet | | |
| | | | Releases android with HTC Hero device; Android Marketplace reaches 5000 applications |
| 2009 | | Maemo 5 injects speed and power into mobile computing; Launches nokia.maemo.org | |
| | Premieres iTunes version 9, New iPod touch, iPod shuffle and Ipod nano | | Android Marketplace reaches 10000 applications |
| | | Accenture's acquisition of Nokia's Symbian Professional Services completed; Nokia announces official Qt port to Maemo 5; Sues apple for infringement of GSM, UMTS and WLAN standards | |
| | Announces Over 100,000 Apps Now Available on the App Store | | Buys AdMob, a mobile advertisement company |
| | | Nokia releases Qt 4.6 | Android Marketplace reaches 20000 applications |
| 2010 | Launches Ipad; App Store Downloads Top Three Billion | Ovi Store delivers content and applications | Unveils Nexus One Phone |
| | | New Ovi Maps with free navigation races past 1 million downloads in a week; Merges software platform with Intel to form MeeGo | |
| | Sues HTC for 20 patent infringements | Skype now available for Nokia smart-phones in Ovi Store | |
| | | Introduces N8 with OS Symbian^3 | |
| | Sells Two Million iPads in Less Than 60 Days | Sues Apple for infringement of 5 patents | |
| | Releases iPhone 4; Sells Three Million iPads in 80 Days | | |
| | | Debuts 'Touch and Type' design with Nokia X3 phone | Android sales outpace iPhone |
| | Apple Introduces New iPod touch, iPod shuffle and Ipod nano. | Appoints Stephen Elop to President and CEO ; Started shipping N8 smart-phone | |
| | | Nokia started shipping C7 smart-phone | Skype becomes available for Android |
| | Apple launches iOS 4.2 for iPad, iPhone & iPod touch; Sues Motorola over Multi touch in | AMD joins MeeGo project | |
| | Android phones | | |
| | | Skype brings video calling to iPhone | |
| | Apple’s App Store Downloads Top 10 Billion. | | Larry Page replaces Eric Schmidt as CEO |
| | Apple Launches Subscriptions on the App Store. | Adopts Windows Phone as primary smart-phone platform | |
| 2011 | Apple Launches iPad 2: Thinner, Lighter & Faster with FaceTime, Smart Covers & 10 Hour Battery; Launches iOS version 4.3 | Sells QT commercial licensing business to Digia | |
| | | Nokia's Ovi Store hits 5 million downloads per day; Announces plans to transfer Symbian software activities to Accenture | Debuts Firefox 4 for Android |

The IT industry is witnessing a war for the first time, which does not involve Microsoft and Nokia in a major role yet. When it comes to mobile device market, the rules of war have changed from the era which saw Microsoft dominating. Today’s mobile platform market is dominated by Apple and Google, with each of them employing a different strategy but more importantly resulting in success for them. Apple with their iPhone device and frequently releasing new versions has certainly got attention of the consumer interest. On the other hand, Google’s Android market share increased by almost 19% from 2009 to 2010, which is an indication of their strategic success (Gartner, 2011). Others players in the market such as Research in Motion undoubtedly control a fair market share but are now facing a new and distinctive type of competition which is more about innovation than surviving on an old success. In the very end, it is all about coming out with a new product which is very useful.

A major aspect of strategy now resides on the open source, the contribution made by developers to promote innovation and richness in platforms and applications. According to Eric Schmidt, executive chairman Google, there are two kinds of players in the industry, the one who makes a very useful, focused product but is a closed competitor and that is Apple. Whereas Google is making technology available to everybody, sharing creativity, and making partnerships is taking a key step towards future innovation. Therefore, with more people involved, more investment will come, and eventually the consumer will choose the open competitor (Zakaria, 2011). However, Apple has been really successful in defining its products for focused groups, but they always try to own and control the working technology. After the success of iPhone 4, the iPad defined a new category of mobile devices. Apple tries to express its vision through its products and their success relies on the fact of not trying to do too much (Huang, 2011). A common success factor for both the companies is the use of open source software but they promote the concept at very different levels, considering the contribution they make in return.

The great Nokia has already closed the Symbian foundation, while open for business it is not an open source anymore. They have also transferred the Symbian development operations to Accenture. A plus point for Symbian Qt developers is that they can port their applications to Android platform from now on, which benefits Google as well. Microsoft's new partnership with Nokia, which makes their Windows 7 the primary platform and buying of Skype, makes their intention quite clear. But what strategy would they employ and whether it will be successful is yet to be seen.

In order to address the third research question on how the third party developers are coping with the announced strategies, a compressive set of software development portals and forums were carefully analyzed according Romano et al. (2003) method for analysis of web-based qualitative data. Free form text communication from different software developers of mobile applications was collected, covering the first quarter of 2011 (January to March) for post classification and analysis. The following Table 5 presents the coding scheme used. A total of 4821 free form text sentences were coded using the dimensions "evaluation", "intention to complement" and "desire for openness".

**Table 5.** Coding scheme used as the reduction step described by Romano et al. (2003)

| Evaluation | Intention to complement | Desire of openness |
|---|---|---|
| Positive | Will not develop for the platform | Developer would like to access platform core source code |
| Negative | Will develop for the platform | Developer would like to be more included in the platform development |
| Neutral | Currently develops for the platform | Developer seeks a more open handset platform for final users. |
| Unknown | Will abandon platform development | Developer seeks to run the platform in a different hardware, other than the one promoted by the vendor. |
| | Neutral | Developer would like tighter control and more filtering efforts from the vendor. |

From the set of collected sentences freely provided from software developers, a considerable number included a perceived evaluation statement from the developers towards the platform. A low number of these sentences, which either used many "jargons" used by software development communities or by being contradictory were coded in the "Unknown" category. It is clearly visible that the Android and Maemo platforms have been given positive remarks by the third party application developers. It is important to point out that many developers continuously provide positive or negative feedback reacting to the platform's continuous development. The discussion is always fomented with the release of new developer tools and development interfaces from the vendors. Surprisingly, there is no evidence, that physical events targeting the software complementers communities such as the Apple Worldwide

Developers Conference, the Google IO, the Symbian's developer's conference, and the Maemo Summit have immediate effects on the evaluation attitude of the platform software developers.

The Fig. 3 below presents the positive and the negative reviews from the developers of different platforms. As visible, the most relevant finding is the low value perception from third party developers on the Symbian platform. Symbian is the only platform with higher negative reviews over positive reviews from developers. However, it is extremely important to notice that experienced developers provide more positive reviews on the Symbian platform as well. Many newcomers confront with the Symbian's recent platform evolutions and leading technical capabilities in power and memory management. The inclusion of the Qt technologies in the Symbian platform captures most of the positive platform reviews.

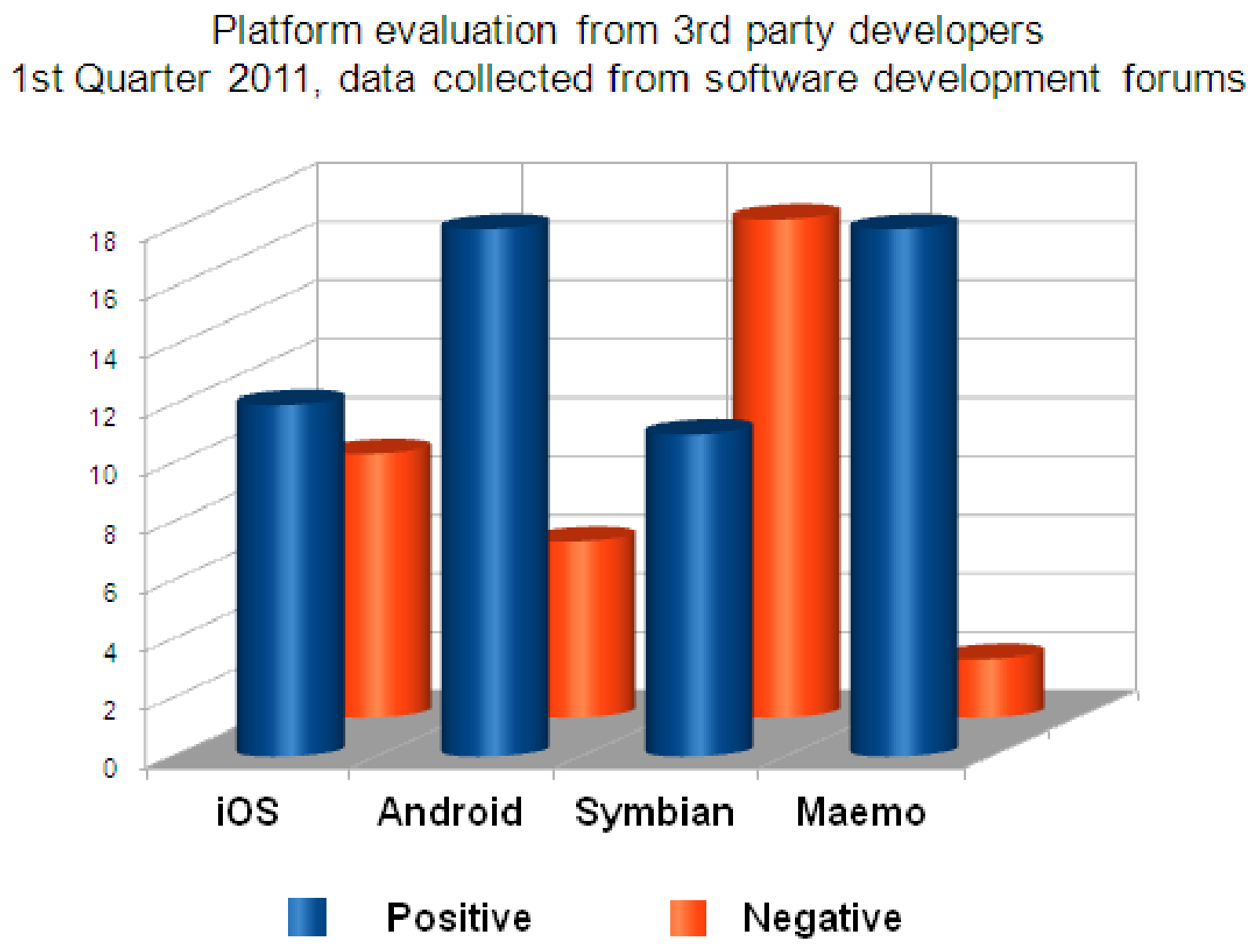

Fig. 3: Developers review by platform

A really high number of the captured sentences were coded regarding the third party developers intentions in complementing the platform, especially on platforms that entered later in the market. The authors noticed that the developers often provide contradictory sentences regarding their “wishes” of complementing the platform; typically a developer provides rich discussion on its first contacts with the platforms. However after several weeks of “coding” a real world application, they change their initial intentions. This means that platform vendors can capture the attention of 3rd party developers, but the platform value is only increased when the developer provides a valuable software package on top of the platform. As visible in Fig. 4 the intentions from Nokia in collecting contributions from third party developers to its Symbian platform seems to be unsuccessful. Many developers expressed their intentions in not adopting the platform and, even more concerning is the high number of developers stating that they will stop developing on the platform. However, it is surprising that many developers are willing to complement to the Maemo platform. Android seems to be the platform in position to collect high contributions from 3rd party software developers. In a market subjected to the forces of network effects, it will bring great value to the Android platform.

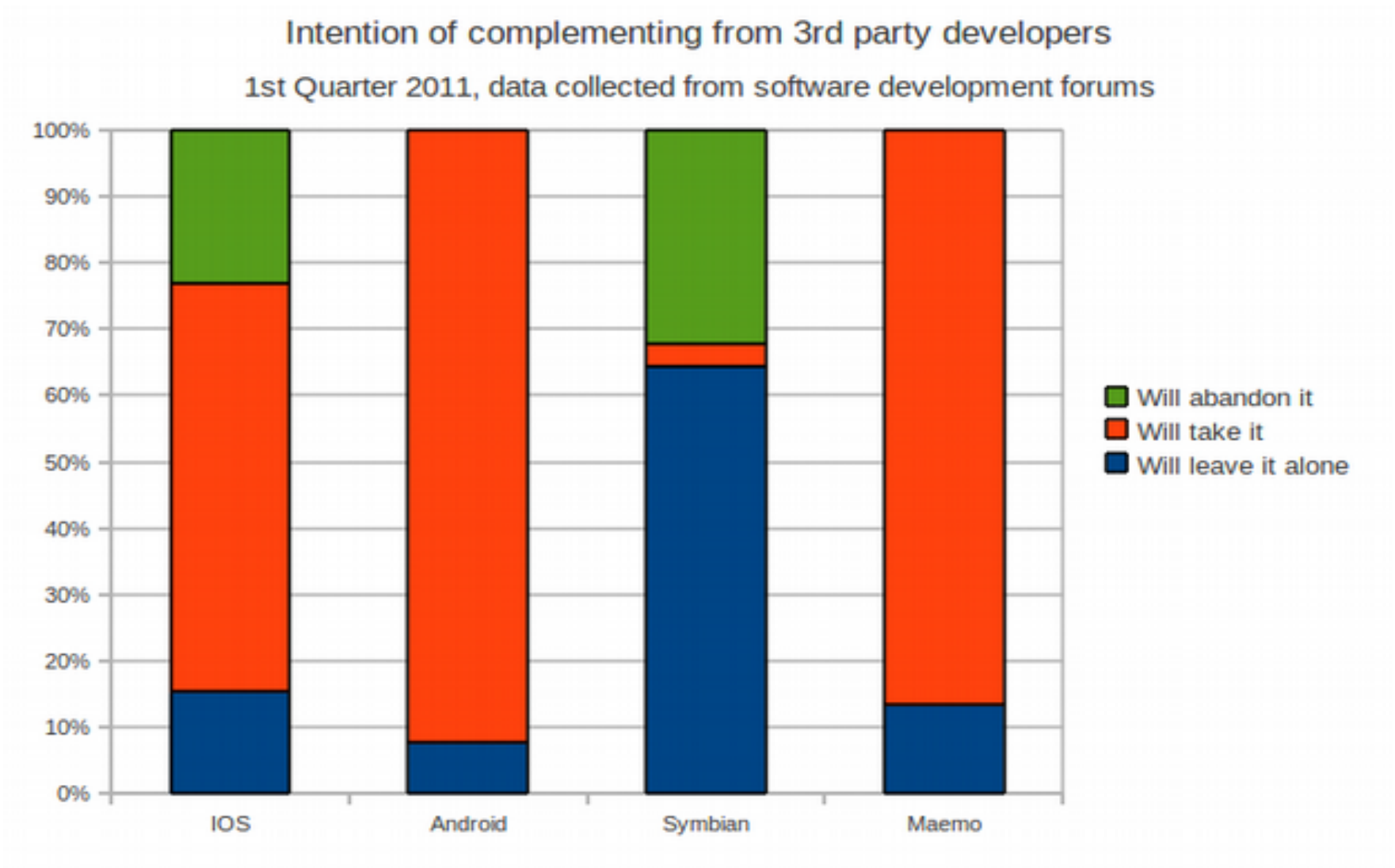

Fig. 4: Developers commitment to each platform

Regarding the platform “openness”, a term referred to many times in the analysis on the Internet sites. It seems that Nokia sponsored platforms raise less debate over its platforms “openness”. The Apple iOS platform, on the other hand, is the one where developers more actively discuss its “openness”, curiously not from a software development point of view but from a critical end-user point of view. If open phones seem to be so important for critical end-users complementing the platform, it would be interesting to perform a future survey on what represents “openness” metric in the normal purchaser decisions. It is important to highlight that no iOS software developer revealed a desire in having the iOS platform running on devices not branded by Apple.

All opinions expressed on “openness” revealed a desire towards a more open-platform, no software developer called for a tighter control and filtering in the iOS ecosystem. The Symbian developers also seem to be very satisfied with the hardware provided by Symbian phone makers such as Nokia and Samsung. As seen in Fig. 5 none of the developers expressed a wish to run the platform on alternative hardware. Until the end of the first quarter of 2011, no Symbian developer had trouble in accessing the platform core source code. Perhaps the most interesting observation is the platform developers call for final-user empowerment. Heavy criticism was given on the impossibility of accessing, by legal means, with “root” privileges in handsets shipped with the iOS and Android platforms. Moreover, the fact that 3rd party applications can be only installed from the vendors Internet markets, commonly referred to as “app stores” or “app markets”, raises strong debate among platform software complementers.

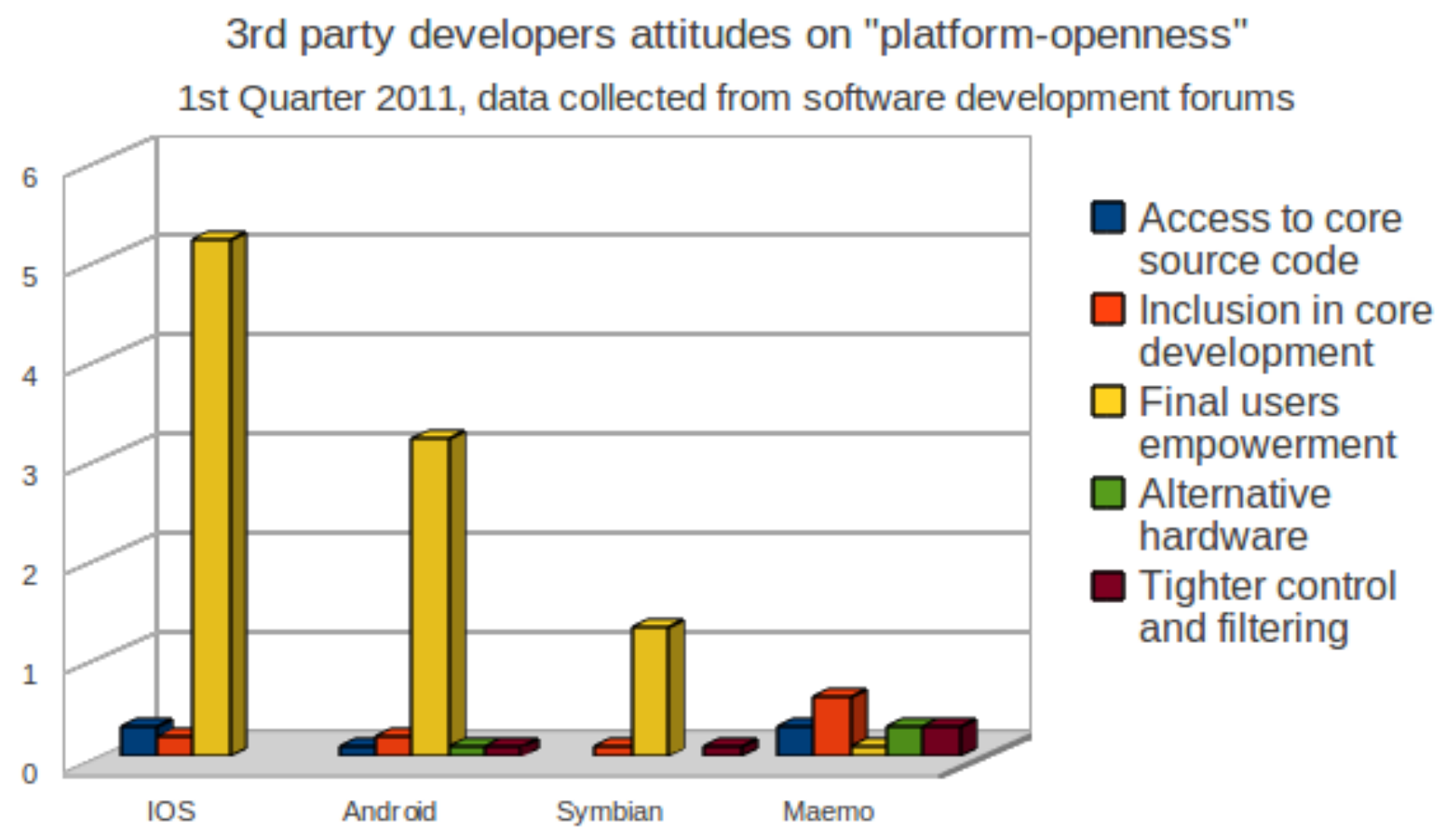

Fig. 5: Developers perceived platform openness

## 4 Limitations and conclusion

Our study encompasses the generalized limitations inherent to case study research. From a set of single cases, the previous reported findings cannot be directly generalized to the current body of theoretical knowledge in Information Systems. However, as pointed by Yin (2002) and Flyvbjerg (2006), case studies play a central role in the academia via generalization as supplement or an alternative to other methods, even if "the force of an example" is still underestimated over formal generalization.

This case study was a great opportunity for testing established knowledge in computer-based platforms from authors such as Anchordoguy (1989), Bresnahan and Greenstein (1999), Hagiu (2004), West (2003), and Gawer and Cusumano (2008). A heterogeneous and evolving definition of what computer-based platforms are; together with an increasing technical sophistication with increasingly complex layers; and an increased networked economy competition embracing phenomenon like open-innovation and co-competition; limit the generalization of the early knowledge built on top of the PC-industry to platform competition in the new mobile-industry. Even if it was not the main goal of this research, this case study can be useful for testing hypothesis gathered from previous literature. Future research should be performed on the concrete testing of some of the previously established research.

It is quite obvious to notice high convergence between the PC and Mobile industries, where vendors try to explore the network effects associated with the compatibility between the different vendor portfolios. As an example, Apple like a traditional PC player is turning into a competitive mobile industry player, Nokia is already selling a lite PC with its netbook product and Google moved completely out of its core by providing a platform impacting the overall telecommunications industry. Vendors are explicitly managing customer lock-in and incompatibility strategies, as described by Shapiro and Varian (1999). One of the most notable example is provided by Ionescu (2009), where Apple continuously disables the access of competitor Palm pre device to its iTunes sync Internet service. On a curious note, Varian is now working as Chief Economist at Google.

Our contributions provide a detailed description on how differently Apple, Goggle and Nokia make use of open-source software components and on how differently they cooperate with the communities of software developers with most of the credits in the integrated technologies. A detailed analytical view on the attitudes from community of software developers on the vendor platforms strategies provides academics and practitioners with valuable data to better understand the mobile-industry landscape.